\documentclass[hyper,a4paper,12pt,oneside]{JHEP3}

\usepackage{amsmath,amscd,epsfig}
\usepackage{amssymb,amsfonts}
\usepackage{graphicx,shortvrb}
\usepackage{newlfont}

\makeatletter
\renewcommand{\subsubsection}{\@startsection{subsubsection}{3}{0mm}{-\baselineskip}{0.5\baselineskip}{\normalfont\normalsize\it}}
\makeatother

\newcommand{\zn}{$\Rf\times\mathbb{C}/{\mathbb{Z}_N}$ }
\newcommand{\SO}[1]{\mathrm{SO}(#1)} \newcommand{\SU}[1]{\mathrm{SU}(#1)}
\newcommand{\U}[1]{\mathrm{U}(#1)}
\newcommand{\alp}{{\alpha'}}

\newcommand{\J}{\mathcal{J}}
\newcommand{\wt}{\widetilde}
\newcommand{\bz}{{\bar z}}

\newcommand{\s}{\sigma}
\newcommand{\idn}{{1\relax{\kern-.35em}1}}
\newcommand{\Cf}{\mathbb{C}}
\newcommand{\Rf}{\mathbb{R}}
\newcommand{\Zf}{\mathbb{Z}}

\newcommand{\csp}[1]{\hspace{#1em},\hspace{#1em}} \newcommand{\hsp}[1]{\hspace{#1em}}
\newcommand{\abs}[1]{\left\vert#1\right\vert}
\newcommand{\bpr}{\bar\partial} \newcommand{\pt}{\partial}
\newcommand{\pdf}[2]{\frac{\partial #1}{\partial #2}}
\newcommand{\vecs}[1]{\vec{#1}^{\;2}}

\DeclareMathOperator{\tr}{Tr} 
\DeclareMathOperator{\re}{Re} \DeclareMathOperator{\im}{Im}
\newcommand{\cno}[1]{\,\textbf{:}#1\textbf{:}\,}
\newcommand{\acno}[1]{\smash{\genfrac{}{}{0pt}{3}{\circ}{\circ}}#1\smash{\genfrac{}{}{0pt}{3}{\circ}{\circ}}}

\newcommand{\bok}[3]{\Bigl\langle #1\Bigl\vert #2\Bigr\vert#3\Bigr\rangle}

\title {Flow of Geometries and Instantons on the Null Orbifold}

\preprint{\hepth{0507067}\\WIS/16/05-JUL-DPP}
\author{M.~Berkooz, Z.~Komargodski, D.~Reichmann, V.~Shpitalnik\\\\
Weizmann Institute of Science,
Rehovot 76100, Israel\\ \\
{\tt E-mail:\\
berkooz, zkomargo, tragula, vshpital@wisemail.weizmann.ac.il} }
\abstract{We study condensation of twisted sector states in the
null orbifold geometry. As the singularity is time-dependent, we
probe it using D-Instantons. We present evidence that the
null-orbifold flows to the $\Zf_N$ orbifold. We also comment on
the subtleties of quantizing the closed superstring in this
background. }


\begin{document}
\setcounter{tocdepth}{2}
\section{Introduction}\label{section1}

The treatment of time-dependent backgrounds and spacelike
singularities remains among the main puzzles of string theory. The
importance of the problem has led to a considerable amount of
work, and some progress has been made using perturbative and
non-perturbative techniques, such as exact CFT's, AdS/CFT etc.
(\cite{Horowitz:1991ap}-\cite{Festuccia:2005pi}). Examples of the
goals of this program include understanding the quantum state at
the big-bang singularity, or at the singularity of the black hole
\cite{Horowitz:2003he}.

In particular, the role of $\alpha'$ corrections vs. perturbative
$g_s$ corrections vs. non-perturbative stringy corrections, at such
singularities, is unclear. It is likely that all are needed in order
to understand the singularity in detail. This is the case in more
familiar stringy singularities, such as the conifold, for which key
effects are understood by quantitative non-perturbative effects, such
as wrapped D-branes becoming light \cite{Strominger:1995cz}, but a
more detailed understanding uses LST or double-scaled LST
\cite{lst}. The latter focuses on a smaller set of degrees of freedom
localized near the singularity (and is a solvable CFT with a varying
$g_s$), which clarifies where and how non-perturbative effects set in,
as well as being computationally useful.

Other effects which are likely to be important are effects beyond
2nd quantized string theory. Already at the level of the
Einstein-Hilbert action generic singularities exhibit a strong
mixing property - the BKL dynamics (for a review of its recent
appearance in supergravity see \cite{Damour:2005mr}). There are no
proposals for a tractable stringy formalism to deal with such
mixing (beyond the low energy effective action, which breaks down
close to the singularity).

We list below some of the motivation for exploring $\alpha'$
effects around the singularity:
\begin{enumerate}
\item In \cite{Fidkowski:2004fc}, a search was carried out for remnants of the BH
  singularity at large N but weak 't-Hooft coupling. No indications of
  the singularity were found. This might suggest that the singularity
  is resolved by $\alpha'$ corrections.

\item $\alpha'$ effects already exhibit interesting and unexpected
  behavior near the singularity. In \cite{Berkooz:2004yy} it was shown
  that already at sphere level a non-commutative geometry like structure
  appears. The latter delocalized the twisted sector states over large
  distances in spacetime. In particular in \cite{Berkooz:2004re} it was shown
  that twisted sector pair creation occurs near the singularity.

\item For the non-rotating extremal BTZ in $AdS_3$, one can construct
  candidates for the microstates of the BH (for a recent review see
  \cite{Mathur:2005zp}). The proposed microstates of the BH are
  solutions of Einstein-Hilbert action, and in particular do not
  require higher $g_s$ corrections. However, they do require $\alpha'$
  corrections in order, for example, to understand the $\Zf_N$
  singularities which occur in some of these configurations.

\item It was recently shown that the singularity and horizon structure
  of a class of supersymmetric black holes changes
  significantly by tree level or 1-loop higher order curvature
  corrections to the effective action \cite{topstr,Sen:2005kj}
\item Although not directly related yet, a problem of singularities in
  causal structure also appears in purely 2D field theory context. The
  Coulomb branch of $\SU2$ (4,4) gauge theory in two dimensions
  \cite{Diaconescu:1997gu} with a single quark flavor has a metric
  which is not positive definite in the IR\footnote{The metric is
    believed to receive only a 1-loop corrections in perturbation
    theory.}. Since the model is the IR of a perfectly well defined
  unitary supersymmetric field theory, this problem has to be resolved
  within the 2D field theory. One usually says that the correct
  degrees of freedom were not identified properly in the IR (already in terms of
  the 2D field theory).

\end{enumerate}

We are interested in mapping what string theory considers to be
small deformations of a singularity, in our case the null orbifold
singularity
\cite{Horowitz:1991ap,Simon:2002ma,Liu:2002ft,Liu:2002kb,Tseytlin:1994sb},
as a step towards understanding its large deformation, which might
be relevant for its resolution. For the null orbifold, we will
explore the relation between it and familiar $\Zf_N$ orbifolds that
posses a mild (and well understood in string theory) time like
singularity.

One can also consider a more detailed role that ``near by
geometries'' can play. Consider for example the relation between
the fuzzball states of  \cite{Mathur:2005zp} and the BTZ geometry.
One way to reconcile the validity of the two descriptions is
examine the amount of mixing the micro-states undergo under any
attempt to probe them. Since they differ from each other only over
small length scales\footnote{We would
  like to thank S. Ross for a discussion of this point.}, they clearly
mix under any such perturbation. This implies that the effective
geometry may not be that of the microstates but something else -
perhaps for some purposes it is the original BTZ black hole
geometry. The situation might be analogous to that of some field
theories which at zero temperature exhibit spontaneous symmetry
breaking, but exhibit symmetry restoration at finite temperature
(above a threshold). The microstates are analogous, in this very rough
analogy, to the true vacua, and the finite temperature minimum in the
origin is analogous to the original black hole singularity, which
dominates the dynamics once mixing is taken into account. In both
cases a complicated enough process, with enough energy, will be
dominated by the symmetric phase (=the background with the black hole
singularity) although most pure states (and in particular the ground
state) are in the broken symmetry phase (which is like the micro-state
description). To go from the symmetric phase to the broken phase one
usually condenses a tachyon (at zero temperature), and hence we would
like to explore the analogues of these tachyons in the case of the
spacelike (or null) singularity.

The null-orbifold singularity has a very concrete relation to the
$\Zf_N$ singularity. Already in \cite{Liu:2002kb} this relation
was touched upon, and we develop it further in the current paper.
In section \ref{section2} we show how precisely the two-cone
null-orbifold is a large N limit of the better understood single
cone $\Cf/\Zf_N$ orbifold (as well as the subtleties associated
with this limit). In section \ref{section3} we discuss the action
of D(-1) branes in this background. In section \ref{section4} we
explore the transition from the null orbifold towards the $\Zf_N$
orbifolds, after condensing an $N$-twisted sector state in the
null-orbifold.

This situation can be summarized in the following diagram
\begin{equation}
\begin{CD}
    \Rf^{1}\times\Cf\big/\Zf_n @>n\rightarrow\infty>\infty\,\text{boosted}>
    \Rf^{1,2}\big/\text{Null}\\
    @V\text{tachyon}V\text{sector }k V     @V\text{tachyon}V\text{sector }mV \\
    \Rf^{1}\times\Cf\big/\Zf_{\frac nk=m} @>n,k\rightarrow\infty\csp{0.5}\frac nk=m>\infty\,\text{boosted}>
    \Rf^{1}\times\Cf\big/\Zf_m
\end{CD}
\end{equation}

The left downward point arrow is the flow of \cite{Adams:2001sv}.
The upper rightward pointing arrow is section 2. The right
downward pointing arrow is section 4 and is the main conclusion of
the paper, which presents evidence that upon condensation of an
$N$ twisted sector mode of the null-orbifold a $\Zf_N$ singularity
appears\footnote{Although in the extreme boost limit, as we will
  discuss later.}.

Section 5 contains a summary and conclusions.

Further details on the $\Zf_N\rightarrow\ null\ orbifold$ are
provided in appendix A. Appendix B quantizes the RNS string on the
null orbifold and shows the emergence of a logarithmic CFT.

As we completed this, two papers appeared which discuss related
models from a different perspective
\cite{Cornalba:2005je,McGreevy:2005ci,Craps:2005wd}

\section{The Null-orbifold and \zn Orbifold}\label{section2}

String theory in orbifolds of the form $(\Rf^{1,2}/\Gamma)\times
\mathcal{C}^\perp$ with $\Gamma$ generating a group isomorphic to
$\Zf$ or $\Zf_N$ were extensively studied. The $\Zf_N$ orbifolds
(where $\Gamma$ is in the elliptic class of $\SO{1,2}$) are well
understood
\cite{Dabholkar:1994ai,Lowe:1994ah,Adams:2001sv,Harvey:2001wm} (for
a review see \cite{Headrick:2004hz}), and the geometry is a cone
perpendicular to the time direction. The null-orbifold studied in
\cite{Simon:2002ma,Liu:2002kb,Tseytlin:1994sb} is generated by
$\Gamma$ in the parabolic class of $\SO{1,2}$, and the geometry
consists of two three dimensional cones with a common tip and a
singular plane crossing the tip.  Unlike the $\Zf_N$ orbifolds the
null-orbifolds is a singular time-dependent
background\footnote{Although it possesses a null
  isometry.}. The singularity at the origin of the cones is still not
completely understood.

In this section we review the quantization of the $\Zf_N$ and the
null orbifold. We introduce the construction of the latter from
the former by a infinite boost. The result we obtain is that the
$\Zf_N$ orbifold converges to the two-cone null orbifold. This is
shown using both the classical geometry and the 1st quantized
string.

\subsection{Classical Results (Geometry)}
To describe orbifolds of  $\mathbb{R}^{1,2}$ we take the
coordinates $x^0, x^1, x^2$ on $\mathbb{R}^{1,2}$, with the flat
metric $ds^2=-d(x^0)^2+d(x^1)^2+d(x^2)^2$ and consider the Killing
vector:
\begin{equation}\label{sec2-Jab}
    J(a,b)=bJ^{02}+aJ^{12}
\end{equation} where
$J^{02}=x^0\partial_2+x^2\partial_0$ is a boost and
$J^{12}=x^1\partial_2-x^2\partial_1$ is a rotation.

For $b<a~$,$J(a,b)$ is in the elliptic class, and conjugate to
$J(\sqrt{a^2-b^2},0)$ using some Lorentz transformation M. The null
orbifold is given by $a=b$.  Choosing $\sqrt{a^2-b^2}=1/N$, the
generator of the null orbifold, $a=b$, is
identified as the limit $~N\rightarrow\infty~$ of a sequence $M_N
J(1/N,0) M_N^{-1}$, where $M_N$ is an N-dependent Lorentz
transformation. This Lorentz transformation is singular when
$N\rightarrow\infty$. It is useful to write the explicit form of
$M_N$.
\begin{gather}
    M_N=\begin{pmatrix}\frac {a}{\sqrt{a^2-b^2}}&0& \frac {b}{\sqrt{a^2-b^2}}\\
                       \frac {b}{\sqrt{a^2-b^2}} & 0 &\frac {a}{\sqrt{a^2-b^2}}\\
                       0 & -1 & 0 \end{pmatrix}
    \cr J(a,b)=M_N J(1/N,0) M_N^{-1}\label{sec2-orb-gen}
\end{gather}
For the pure rotation case ($b=0$) we will use the familiar convention:
\begin{equation}
    \left(
      x^0 ~,~   z ~,~ \bar z\right)\cong
    \left(
      x^0 ~,~  e^{\frac{2\pi i}{N}}z ~,~  e^{-\frac{2\pi i}{N}}\bar z \right)
      \qquad\text{with}\quad z=\frac{x^1+ix^2}{\sqrt2}.
\end{equation}
The generator of the null Orbifold twisting operator acts on the
light cone coordinates
\begin{equation}
  x^+=\frac {x^0-x^1}{\sqrt{2}} \ \ \ \  x=x^2 \ \ \ \ \
  x^-=\frac{x^0+x^1}{\sqrt{2}}
\end{equation}
by
\begin{equation}
    J^{null} = a \begin{pmatrix}  0 & 0 & 0 \\ \sqrt{2} & 0 & 0\\0 & \sqrt{2} &    0
\end{pmatrix}
\end{equation}

We shall follow the convention of \cite{Liu:2002kb} in defining the
null orbifold by choosing $a=\frac{\nu}{2\pi \sqrt{2}}$ and the
identification takes the following form
\begin{align}
&x^+\thicksim x^+ \cr & x\thicksim x+ \nu x^+ \cr & x^-\thicksim
x^-+\nu x+\frac{\nu^2}{2}x^+
\end{align}

A fundamental domain of the space looks like two 3 dimensional cones
emanating from $x^+=x=0$, one towards $x^+>0$ and one towards $x^+<0$,
and two codimension 1 cones in the plane $x^+=0$  pinching at
$x=x^-=0$. Any two null orbifolds (differing by the value of $\nu$)
are related by boosts in the $x^1$ direction.

We show how by the procedure described above, of boosting by
$M_N$, we may actually understand geometrically that there is a
singular limit which relates the $\Zf_N$ space with the null
orbifold space. This isn't straightforward as the $\Zf_N$ is an
one cone space  and the null orbifold has two cones (that are not
contained in the singular plane). Indeed for any finite $N$, we
remain with an one cone fundamental domain. Some of the orbits in
the $x^+,x,x^-$ coordinates for large N are plotted in figure
\ref{fig:orbits}. One confirms the impression from the figure,
that the orbits become localized around a fixed $x^+$ with a
spread in $x^+$ which is $\frac1N$ that of $x^-$. At
$N\rightarrow\infty$ we observe that the slopes go to infinity. In
this limit the orbits are contained in the $x \ x^-$ plane, at
fixed $x^+$. The infinity slopes orbits are exactly the parabolas
of the null orbifold. Hence the single cone orbifold maps onto the
two cone null-orbifold geometry.
\\
\FIGURE{
    \centerline
    {\vbox{\hspace{-0.7cm}
    \includegraphics[width=0.45\textwidth]{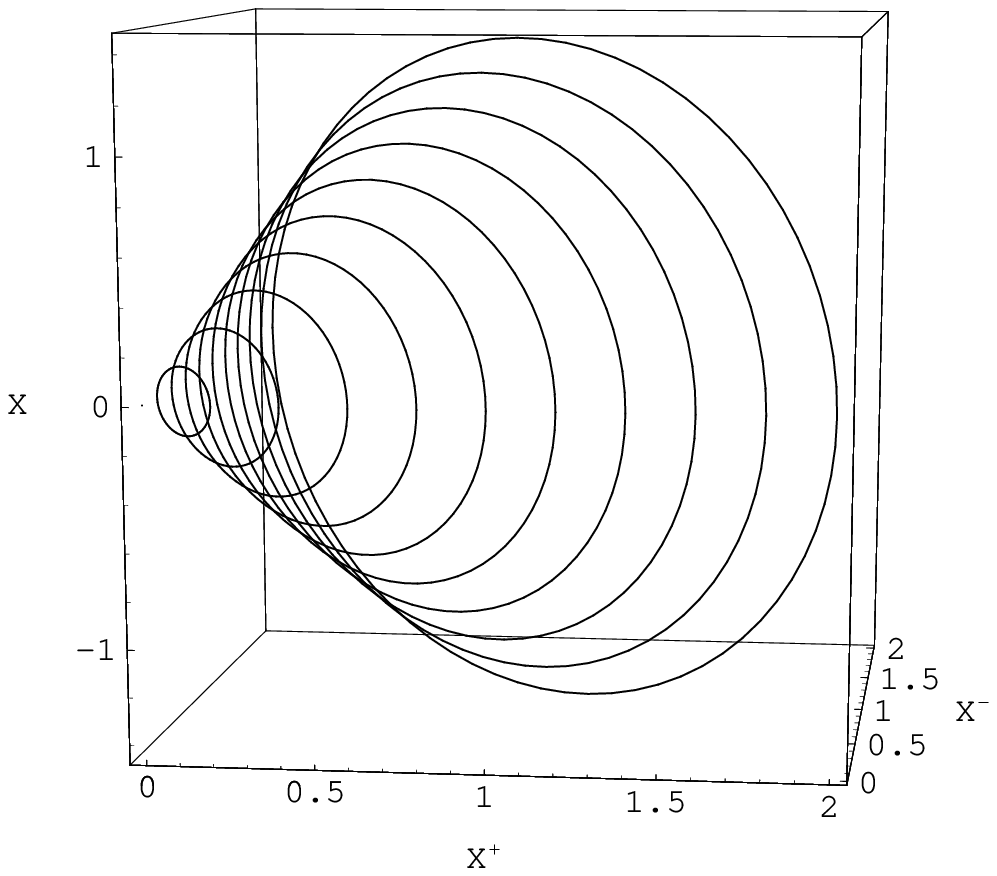}\hspace{.8cm}
    \includegraphics[width=0.48\textwidth]{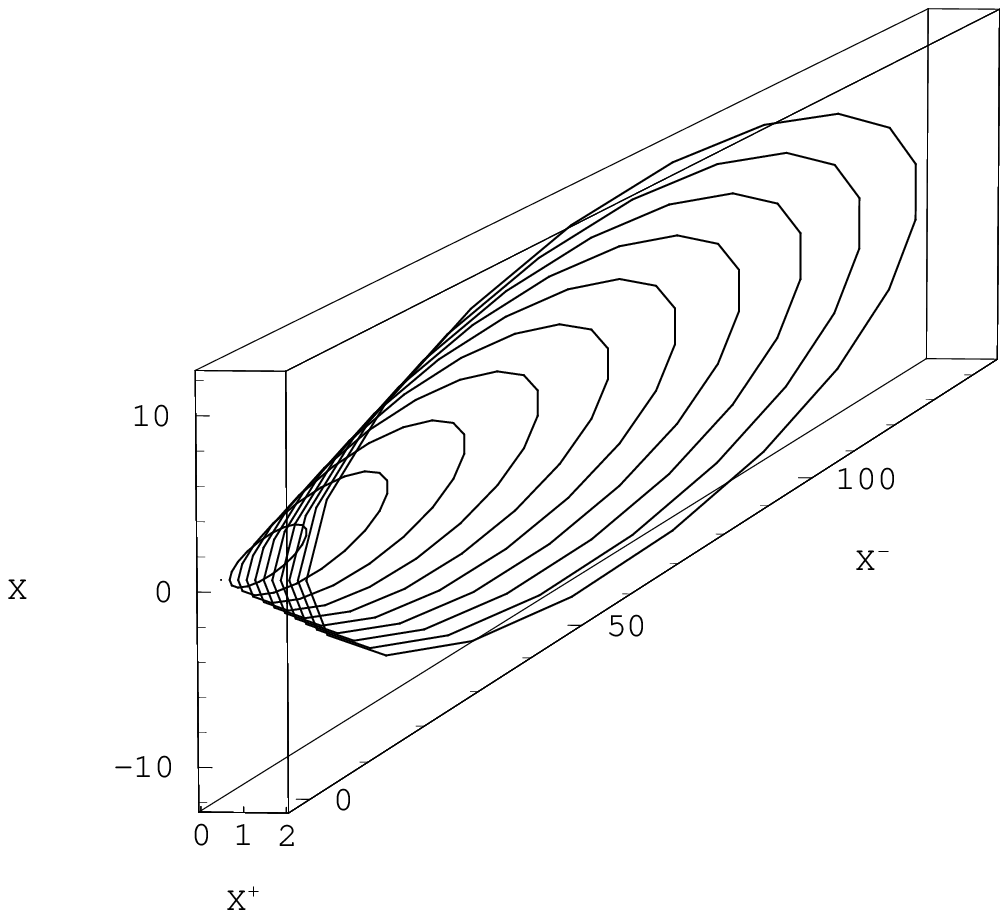}}}
    \caption{Particular orbits of the boosted cone for N=3 (left) and N=6 (right). The orbits are generated from
    points on the $x^+$ axis in the segment $x^+\in(0,2)$ by a continuous
    action of the orbifold generator $J(a,b)$. For large N the
    orbits are stretched in the $x-x^-$ plane, such that at the
    limit $N\rightarrow \infty$ they are confined to that plane
    (i.e $x^+$ is constant within a single orbit). For the limit $N\rightarrow
    \infty$ we expect the orbits to close only at infinity, thus
    dividing the single cone to a double cone.
\label{fig:orbits}} }
%

\subsection{Hilbert space: Untwisted sector}\label{sec2-subsec-untwisted}

As is usual in orbifolds, the untwisted wave functions are projections
of the wave functions of the covering space. The latter are plane
waves on $\Rf^{1,2}$, and the untwisted sector of the orbifold
(focusing on scalar functions) is \cite{Liu:2002kb}
\begin{equation}\label{sec2-cov-id}
    \Psi^{orb.}_{k,s}=\int^{\infty}_{-\infty}ds\,e^{2\pi
    s\left(il+\hat J\right)}\psi_k(x)
    \csp1l\in\mathbb{Z}
\end{equation}
where ${\hat J}$ is the action of the null boost generator on the
function $\psi_k$ (the wave function in flat space).

The formula for the \zn orbifold is similar. The invariant wave
functions of the elliptic orbifold are
\begin{gather}\label{sec2-zn-int}
    \Psi_{k,l}^{N}=\frac{N}{2\pi}e^{-ik^0x^0}\oint^{2\pi}_{0} d\theta e^{ik\bar{z}e^{i\theta}+i\bar{k}ze^{-i\theta}+iN\theta l}
    \csp1l\in\mathbb{Z}\cr
    \text{with,}\qquad k=\frac{k^1+ik^2}{\sqrt{2}}\hsp2 z=\frac{x^1+ix^2}{\sqrt{2}}
\end{gather}
These integrals can be evaluated in terms of Bessel functions:
\begin{equation}\label{sec2-zn-wf}
    \Psi_{k,l}^{N}=N e^{-ik^0x^0+iN\phi
    l}J_{Nl}(2u)\csp1
    u{e^{i\phi}}\equiv-i\bar kz
\end{equation}
It is easy to verify the completeness of this basis. We also choose
$k$ to be
real.

For
the null-orbifold the integration in \eqref{sec2-cov-id} is Gaussian
(The operator $\hat J^{null}$ is nilpotent of order 3), and the
wave-function matches the results of \cite{Liu:2002kb} (after fixing
the phase and taking $\nu=2\pi$ until the end of the section):
\begin{gather}\label{sec2-null-wf}
    \Psi^{null}_{k,l}
    =\frac{\exp\left[i\left(-k^+x^--k^-x^++\frac{(l-k^+x^2)^2}{2k^+x^+}\right)\right]}{\sqrt{2\pi}\sqrt{ik^+x^+}}\csp1l\in\Zf\cr
    \text{where}\qquad k^\pm=\frac{1}{\sqrt2}(k^0\mp k^1)\hsp2 x^\pm=\frac{1}{\sqrt{2}}(x^0\mp
x^1)
\end{gather}
This is the wave function on the three dimensional cones. On the
singular co-dimension 1 cones, it is a distribution.

We have shown that one can take the limit of the geometry.
However, since we are interested in a CFT statement, it is more
meaningful to show that the limit of the set of wave functions on
the single cone $R^2/Z_N$ is the set of all wave functions on the
null orbifold.
 Under an $M_N$ boost, the wave-function transforms as:
\begin{align}
    \Psi^{(a,b)}_{k,l}(x)\equiv &\int^{\infty}_{-\infty}ds\,e^{2\pi
    s\left(il+\hat J(a,b)\right)}\psi_k(x)=\cr
    \propto &\int^{\infty}_{-\infty}ds\,\hat{M_N}e^{2\pi
    s\left(il+\hat J(\frac1N,0)\right)}\hat{M_N^{-1}}\psi_k(x)
    \propto \Psi^N_{\widetilde{k},l}(M^{-1}_Nx)
\end{align}
where $\widetilde{k}_\mu=k_\nu{(M_N)^\nu}_{\mu}$. In the limit
$N\rightarrow\infty$, after normalizing the wave functions, we expect
to find that:
\begin{equation}\label{sec2-wf-limit}
    \Psi^{Null}_{k,l}(x)=\lim_{N\rightarrow\infty}\Psi^{N}_{kM_N,l}(M_N^{-1}x).
\end{equation}
To demonstrate that this limit is well defined (and not just
formal) we will show it explicitly using the wave-function
\eqref{sec2-zn-wf} and \eqref{sec2-null-wf}. The limit is defined
so that the parameters $a,b$ in $M_N$ approach their final
(common) value symmetrically\footnote{This is a necessary
requirement. For other prescriptions we have not been able to show
that one obtains the null orbifold wave functions.} keeping $a>b$
\begin{align}\label{sec2-symmetric}
  a=\frac{1}{\sqrt{2}}+\frac{1}{2\sqrt{2}}N^{-2}+O(N^{-4}) \hsp4
  b=\frac{1}{\sqrt{2}}-\frac{1}{2\sqrt{2}}N^{-2}+O(N^{-4})
\end{align}
Applying this transformation to the wave-functions
\eqref{sec2-zn-wf} in the large N limit we find (further details are
provided in appendix \ref{appendix a}):
\begin{align}\label{sec2-result of boost}
   \Psi^{(a,b)}_{k;\,l}(x)=&
    \frac{(\text{sign}(k^+x^+))^{Nl}}{\sqrt{2\pi ix^+k^+}}
    \Biggl\{\cr
    &\exp\biggl[i\frac{\left(l-x^2k^++x^+k^2\right)^2}{2x^+k^+}+ik^2x^2-ik^+x^--ik^-x^++O(N^{-1})\biggr]+\cr
    &+(-)^{Nl+1}\exp\biggl[-i\frac{\left(l+x^2k^+
    -x^+k^2\right)^2}{2x^+k^+}-ik^2x^2-i2N^2k^+x^++O(N^{-1})\biggr]
    \Biggr\}\cr
\end{align}
$k^2$ may be set to zero using a Lorentz transformation. Taking the
limit $N\rightarrow\infty$ the second term is zero\footnote{To see
  that, it should be considered as a distribution: any integral with a
  well behaved function vanishes in the limit. The exact statement is
  explained in appendix \ref{appendix a}.} and the first term reduces
exactly to the null-orbifold wave function\footnote{The sign in front
  is irrelevant as $Nl$ can be chosen even throughout.}
\eqref{sec2-null-wf}, thus completing our proof.

\subsection{Hilbert space: Twisted
sectors}\label{sec2-subsec-twisted} The authors of
\cite{Liu:2002kb} showed that the zero mode wave functions in the
twisted sectors of the null-orbifold are (in our conventions of
normalizing the wave functions):
\begin{equation}\label{sec2-null-twisted}
    \Psi^w_{m,p^+,J}=\sqrt\frac{1}{2\pi ip^+x^+}\exp\left[-ip^+x^--i\frac {m^2}{2p^+}x^++i\frac{p^+}{2x^+}\left(x+\frac {J} {p^+}\right)^2-i\frac{w^2(x^+)^3}{6(\alp)^2p^+}\right]
\end{equation}
with $J\in \mathbb{Z}$ and $m$ is the three dimensional mass given
by the on shell condition:
\begin{equation*}
    2p^+p^-=m^2=-\frac 4 {\alpha '}+\vecs{p}_{\bot}
\end{equation*}
One observes that this wave function is the invariant combination
of wave functions describing particle in the presence of an
extremal configuration of electric-magnetic fields $E=\pm B$ (in
analogy to \cite{Pioline:2003bs}). The absolute value of this
field is proportional to the twist parameter.

We now write the twisted sector wave functions of the \zn orbifold
in the same manner by considering a particle in the presence of a
magnetic field\footnote{For completeness, we derive this idea on
general orbifold of $\mathbb{R}^{1,2}$ in the next subsection.}.
The most convenient covering space wave functions are radial
strips.
\begin{equation}\label{sec2-radial strip}
\Psi^{k/N}_{p^0,j,n_r}=e^{ij\phi}e^{-ip^0x^0}R_{\abs{j},n_r}\left(\frac{k}{N\alp}r^2\right)
\end{equation}
Where $r,\phi$ are the usual polar coordinates on the plane,
$j\in\mathbb{Z}$ and the radial function is given by
\begin{equation}\label{sec2-radial}
 R_{\abs{j},n_r}(\xi)=e^{-\xi/2}\xi^{\abs{j}/2}F\bigl(-n_r,\abs{j}+1,\xi \bigr)
\end{equation}
$F$ is the degenerate (confluent) hypergeometric function. The
orbifold invariance condition is simple in these coordinates and is
given by $j\in N\mathbb{Z}$. The wave function describes a particle
with mean distance of $\sim n_r\sqrt{\alp}$ from the origin and the
wave function has  $\sim j$ oscillations. The conformal weight
(energy) of this state is
\begin{equation}\label{sec2-energy}
    E=-\frac{\alp}{4}(p^0)^2+\frac kN\bigl(n_r+\frac12(\abs{j}-j+1)\bigr)
\end{equation}

The relation to the usual states of the CFT, which are given by
acting with creation quasi zero modes, is the following :
\begin{equation}\label{sec2-relation CFT}
L_0=E \hsp2 \frac kNj=L_0-\tilde{L}_0
\end{equation}
Using these relations one may obtain the CFT meaning of $n_r$
\begin{equation}\label{sec2-n_r} \frac kN n_r = \frac
  \alp 4
  (p_0)^2+L_0-\frac12\left(\abs{L_0-\tilde{L}_0}-(L_0-\tilde{L}_0)+1\right)
\end{equation}

In order to boost the wave function we boost the coordinates as in
\eqref{sec2-wf-limit} but the quantum numbers are more subtle. In
order to match the quantum numbers of the boosted wave function with
the quantum numbers of the null orbifold it is very suggestive to
use the geometric interpretation of $n_r$. Indeed we expect it to
transform as $\sim(p^1)^2+(p^2)^2$. The exact way is inferred from
\eqref{sec2-energy} and we simply replace $n_r$ everywhere by
\begin{equation}
    n_r\equiv \frac{\alp N}{4k}((p^1)^2+(p^2)^2).
\end{equation}
The quantum number $j/N$ (which is integer) will be mapped exactly
to $l$ as can already be seen from the fact the azimuthal part of
the wave function \eqref{sec2-radial strip} coincides with the
untwisted wave function azimuthal part \eqref{sec2-zn-wf}.

To recapitulate, the wave function of the twisted \zn which is
ready to be boosted with appropriately chosen quantum numbers
\begin{equation}\label{sec2t-whittaker-R}
    \Psi^{k,N}_{p^0,\abs{p},l}=
    e^{iNl\phi}e^{-ip^0x^0}R_{\abs{Nl},\frac{\alp
    N}{4k}\abs{p}^2}\left(\frac{k}{N\alp}r^2\right)
\end{equation}
In doing the boost it is convenient to move to Whittaker functions
which are related to $R$ according to
\begin{equation}\label{sec2-whittaker-R}
    M_{\frac{\abs{j}}2+n_r+\frac12,\frac{\abs{j}}2}(z)=z^{1/2}R_{\abs{j},n_r}
\end{equation}
Using the $M_N$ above, and the asymptotics of the Whittaker function
one can show that \eqref{sec2t-whittaker-R} converges to \eqref{sec2-null-twisted}.

\subsection{First Quantization of the string}
Applying the quantization scheme introduced in
\cite{Pioline:2003bs} we quantize the bosonic string on
$(\Rf^{1,2}/\Gamma)\times \Rf^{d-3}_\perp$. A more detailed
discussion including quantization of the superstring is postponed
to appendix \ref{appendix b}.

The worldsheet action and monodromies are:

\begin{align}
    &S=\frac1{4\pi\alp}\int_{-\infty}^\infty d\tau\int_0^{2\pi}d\s~
    \eta_{\mu\nu}\left(\pt_\tau X^\mu\pt_\tau X^\nu-\pt_\s X^\mu\pt_\s
    X^\nu\right)\cr
    &X(\s+2\pi,\tau)=e^{2\pi w \J}X(\s,\tau)
\end{align}
where $~w\in\Zf~$ is the twisted sector number and $\J$ is matrix
defined from the differential realization of \eqref{sec2-Jab}:
\begin{equation*}
    \hat J(a,b)=X^{\mu}{\J_\mu}^{\nu}\pdf{}{X^\nu}
\end{equation*}
The mode expansion in a twisted sector $w\neq0$:
\begin{multline}
    X^\mu(\s,\tau)=
    {\left[e^{w\J\s}\right]_{\nu}}^{\mu}{X_z}^\nu(x,p\,;\tau)+
    i\sqrt{\frac\alp2}\sum_{n\neq0}{\left[\frac{e^{-i(n+iw\J)(\s+\tau)}}{n+iw\J}\right]_{\nu}}^{\mu}\alpha_{n}^\nu+\\
    +i\sqrt{\frac\alp2}\sum_{n\neq0}{\left[\frac{e^{i(n-iw\J)(\s-\tau)}}{n-iw\J}\right]_{\nu}}^{\mu}\tilde\alpha_{n}^\nu
\end{multline}
With the zero-mode
\begin{gather}
    \text{Null:}\hsp3{X_z}^\mu(x,p\,;\tau)
    ={{\cosh\bigl(w\tau
    \J\bigr)}^\mu}_\nu x^\mu+{{\Bigl[\idn+\frac12\cosh\bigl(w\tau \J\bigr)\Bigr]}_\nu}^\mu\,\frac{2\alp\tau}{3}
    p^\nu\cr
    \text{Other:}\hsp3{X_z}^\mu(x,p\,;\tau)
    ={{\cosh\bigl(w\tau
    \J\bigr)}^\mu}_\nu x^\mu+{\left[(w\J)^{-1}\sinh(w\J\tau)\right]_\nu}^\mu\,\alp p^\nu
\end{gather}
Where $x^\mu$ and $p^\mu$ satisfy the canonical commutation
relations $[x^\mu,p^\nu]=i\eta^{\mu\nu}$.\\ Introduce the
operators:
\begin{equation}
    \alpha_0=\sqrt{\frac1{2\alp}}(w\J x+\alp p)\hsp2    \tilde\alpha_0=\sqrt{\frac1{2\alp}}(w\J x-\alp  p)
\end{equation}
Although assigned subscript $0$, these aren't real zero modes in
general and may possess nonzero conformal weight as we discuss
below. Hence, we will properly name them quasi zero modes. The
commutation relations between the modes:
\begin{equation}
    [\alpha_n,\alpha_m]=
    \delta_{n+m}(n+iw\J)\,\eta\csp1
    [\tilde\alpha_n,\tilde\alpha_m]=\delta_{n+m}(n-iw\J)\,\eta\csp1
    [\alpha_n,\tilde\alpha_m]=0
\end{equation}
The Virasoro generators are
\begin{equation}
    L_n =-\frac1{\alp}\oint\frac{dz}{2\pi i}z^{n+1}\cno{\pt X\pt
    X(z)}=\frac12\sum_{m}\acno{\alpha_m\eta\alpha_{n-m}}+\delta_{n,0}a^X(w)
\end{equation}
For any choice of normal ordering scheme of the quasi zero modes
the zero point energy in the bosonic sector is
\begin{equation}
    a^X(w)=
    \frac{w^2}4\tr(\J^2)+\bok{vac}{\frac12\eta_{\mu\rho}\alpha_{0}^\mu\alpha_{0}^\rho}{vac}_{w}
\end{equation}
The correct choice of normal-ordering depends on the $\SO{1,2}$
class of the orbifold identification generator:
\begin{itemize}
    \item Elliptic class: the quasi zero modes have positive and negative conformal
    weights. Therefore, we naturally choose the positive conformal weight to
    annihilate the twisted vacuum state. The collection of states
    generated by acting with the creation operators is the Hilbert
    space of a (1st quantized) particle in a uniform magnetic field.
    \item Hyperbolic class: the quasi zero modes have
    pure imaginary conformal weight, as discussed in
    \cite{Pioline:2003bs} we should choose the reality conditions to be consistent with the commutation relations. This way, $L_0$ and $\tilde L_0$ are
    manifestly hermitian. The result
    is the Hilbert space of a particle in uniform electric field.
  \item Parabolic class: the quasi zero modes have conformal weight
    zero. The Hilbert space is of a particle in electric-magnetic
    fields which are equal in magnitude. One choice of quantization is
    that of \eqref{sec2-null-twisted}.
\end{itemize}

\section{D(-1) Instantons probes}\label{section3}
\subsection{The World-Volume Theory}
We wish to examine the null orbifold deformed by some tachyon
condensate. We shall probe the space with instantons (D(-1)
branes), following the discussion of
\cite{Adams:2001sv,Taylor:1996ik}. In this section we
develop the technology needed, beginning with the non deformed null
orbifold. The open string theory on the instantons is a matrix
theory of the collective coordinates (fermionic and bosonic). We
focus on the bosonic degrees of freedom, they are parameterized by
the covering space coordinates:
\begin{equation*}
    (X^+,X^-,X)\in\Rf^{2,1}\csp1 (Y^3,Y^4,\ldots
    Y^9)\in\Rf_\perp^7
\end{equation*}
Under the projection each D(-1) instanton has infinitely many
images, we use Chan-Paton indices that span the adjoint of
$\U\infty$ in the covering space. The null-orbifold projection
should break the $U(\infty)$ to $U(1)^\infty$, even at the
singularity. We use the orbifold projection:
\begin{align}\label{sec3-recursion}
    &Y^a_{i,j}= Y^a_{i-1,j-1} \csp4 a=3...9\cr
    &X^+_{i,j}= X^+_{i-1,j-1} \cr
    &X_{i,j}= X_{i-1,j-1} + \nu X^+_{i-1,j-1} \cr
    &X^-_{i,j}= X^-_{i-1,j-1} + \nu X_{i-1,j-1}+\frac{1}{2}\nu^2X^+_{i-1,j-1}
\end{align}
These recursive equations are linear and can be easily solved. The
solution can be neatly written using the following matrices (which
also define an infinite closed algebra):
\begin{gather}
    (\beta^m_l)_{i,j}\equiv \frac{(i+j)^l}{2^l}\delta_{i,j-m}\cr
    [\beta^m_l,\beta^{m'}_{l'}]
    =2\sum_{p=0}^l\sum_{p'=0}^{l'}\left(\frac m2\right)^{l'-p'}\left(-\frac {m'}2\right)^{l-p}
    \beta_{p+p'}^{m+m'}\,\delta_{p'-p\in2\Zf+l'-l+1}\,\binom{l}{p}\binom{l'}{p'}
\end{gather}
The collective coordinates (bosonic) fields which solve
\eqref{sec3-recursion}:
\begin{align}\label{sec3-modes}
    &Y^a= \sum_{m\in\mathbb{Z}}y^a_m\beta^m_0&
    &X=\sum_{m\in\mathbb{Z}}x_m\beta^m_0+\nu x_m^+\beta_1^{m}\cr
    &X^+=\sum_{m\in\mathbb{Z}}x^+_m\beta^m_0&
    &X^-=\sum_{m\in\mathbb{Z}}x^-_m\beta^m_0+\nu
    x_m\beta_1^{m}+\frac{\nu^2}2 x_m^+\beta_2^{m}
\end{align}
along with the reality conditions
\begin{equation*}
    (y^a_m)^*=y^a_{-m}\hsp2
    (x^+_m)^*=x^+_{-m}\hsp2
    (x^-_m)^*=x^-_{-m}\hsp2
    (x_m)^*=x_{-m}.
  \end{equation*}

The world volume low energy effective Lagrangian of the instantons
may be obtained by dimensionally reducing the 10d Super Yang Mills. The action on the world-volume
is\footnote{The normalization factor $Z_0$ (which is simply
$\sum_{-\infty}^\infty 1$) is set to remove an overall infinite
factor coming from the traces.}:
\begin{align}
    \mathrm{S}=&-\frac1{2Z_0}\sum_{\mu,\nu=0}^{9}\tr\left([X^\mu,X^\nu][X_\mu,X_\nu]\right)=\cr
    =&\sum_{m+n+m'+n'=0}\biggl[-y^a_my^a_{m'}x^+_nx^+_{n'}m
    m'+ x_nx^+_m x_{n'}x^+_{m'}(m m'-2nm')+\cr
    &\hsp8
    +2 x^+_mx^-_n x^+_{m'}x^+_{n'}nm'+\nu^2x_m^+x_n^+
    x^+_{m'}x^+_{n'}\frac{mm'n^2}4\biggr]
\end{align}
It is useful to represent the above action using real bosonic
fields living on $S^1$, identifying the modes as Fourier
coefficients for a real field on $S^1$ as in \cite{Taylor:1996ik}
:
\begin{equation}\label{sec3-fields-sig}
    u_m=\int_0^{2\pi}\frac{d\sigma}{\sqrt{2\pi}}\,
    U(\sigma)e^{-im\sigma}\csp2
    U=\left(X^+,X^-,X,Y^3,\ldots Y^9\right)
\end{equation}
with the action:
\begin{equation}\label{sec3-sig-action}
    \mathrm{S}
    =\int^{2\pi}_0\frac{d\sigma}{2\pi}\biggl[
    (X^+\dot{Y^a})^2-(X\dot{X^+})^2+2\dot{X}\dot{X}^+XX^+
    -2(X^+)^2\dot{X}^+\dot{X}^--\frac{\nu^2}{12}(\dot{X}^+)^4\biggr]
\end{equation}
One may think of this action as the (analog of) T dual action, for D0
branes wrapping a space-like cycles. We show in the next subsection
how we may identify the gauge fields and gauge invariant operators of
the theory.
\subsection{Symmetries} The action of the $U(1)_m$ factor in the
gauge group on the fields \eqref{sec3-modes} is generated by
$\beta^m_0$, inducing the following transformation law for the modes:
\begin{align}
    &e^{\alpha Q_m}y_n^a=y_n^a&
    &e^{\alpha Q_m}x_n=x_n+\alpha\nu mx_{n-m}^+\cr
    &e^{\alpha Q_m}x_n^+=x_n^+&
    &e^{\alpha Q_m}x_n^-=x_n^-+\alpha\nu
    mx_{n-m}+\frac{1}{2}(\alpha\nu m)^2x_{n-2m}^+
\end{align}
Although it is a $0+0$ model, one still considers the above
transformations, which are symmetries of the Lagrangian, as gauge
transformation. Using the $\sigma$ representation we can combine the
transformation into a single gauged $U(1)$ and an associated arbitrary
$\Lambda(\sigma)$ living on the $S^1$ which is the gauge
transformation $0$-form:
\begin{align}\label{sec3-gauge-sym}
    &{Y^a}{\,'}=Y^a&
    &{X  }{\,'}=X+\nu\,\partial_\sigma\Lambda\,X^+\cr
    &{X^+}{\,'}=X^+&
    &{X^-}{\,'}=X^-+\nu\,\partial_\sigma\Lambda\,X
    +\frac{\nu^2}{2}\left(\partial_\sigma\Lambda\right)^2\,X^+
\end{align}
The gauge transformations above are all connected to the identity
of $U(1)^\infty$, however there is another element in the group
which is disconnected from the identity (we will refer to it as
'large gauge transformation'). It's action on the fields
\eqref{sec3-modes}:
\begin{equation*}
    U'_{ij}=\sum_{kl}(\beta^0_1)^{-1}_{ik} U_{kl} (\beta^0_1)_{lj}
\end{equation*}
The transformations of this element on $\sigma$-representation
fields \eqref{sec3-fields-sig}:
\begin{align}\label{sec3-gauge-large}
    &{Y^a}{\,'}=Y^a&
    &{X  }{\,'}=X+\nu\,X^+\cr
    &{X^+}{\,'}=X^+&
    &{X^-}{\,'}=X^-+\nu\,X
    +\frac{\nu^2}{2}\,X^+
\end{align}
These transformations (and all successive transformations
generated by it) can be viewed as a modification on
\eqref{sec3-gauge-sym} by allowing a specific non-periodic
boundary condition for the $0$-form $\Lambda(\s)$.

The action \eqref{sec3-sig-action} is also invariant under
translations in $\sigma$, this is the quantum $\U1$ symmetry which
insures the conservation of winding number of the string and the
momentum in the T dual picture.

\subsection{Moduli space}

As a check of the formalism we will verify that the classical moduli
space becomes the position of a single instanton in the fundamental
domain of the null orbifold. The equations of motion derived are:
\begin{subequations}
\begin{align}
    \label{sec3-eomy}
    & \frac{d}{d\sigma}\left[(X^+)^2\,\frac{d Y^a}{d\sigma}\right]=0\\
    \label{sec3-eomxm}
    & \frac{d}{d\sigma}\left[(X^+)^2\,\frac{d X^+}{d\sigma}\right]=0\\
    \label{sec3-eomx}
    & \left[\frac{d^2(X^+)^2}{d\sigma^2}-\left(\frac{d}{d\sigma} X^+\right)^2\right]X=0\\
    \label{sec3-eomxp}
    & X^+\left[X^+\frac{d^2X^-}{d\sigma^2}-\frac12\frac{d^2(X^2)}{d\sigma^2}+\sum_a (\dot Y^a)^2\right]
    +\frac{d}{d\sigma}\left[X^2\frac{dX^+}{d\sigma}
    +\frac{\nu^2}{6}\left(\frac{dX^+}{d\sigma}\right)^3\right]=0
\end{align}
\end{subequations}
Solving the second and third equations we see that $X^+=const$. By
gauge invariance we can make $X$ constant (for a nonzero $X^+$).
Then the equations of motion and periodicity constraint for $X^-$ and
$Y$ to be constants. The large gauge transformation
\eqref{sec3-gauge-large} is still not fixed, so the moduli space
is:
\begin{equation}
    \mathcal{M}=\left\{X^+,X^-,X,Y^3,\ldots
    Y^9\right\}\biggm/\left\{\begin{array}{l}
      X\cong X+\nu X^+ \\
      X^-\cong X^-+\nu X+\frac{\nu^2}2 X^+ \\
    \end{array}\right\}
\end{equation}
The above solution is the Higgs branch and as expected it is the
null-orbifold.

The analog of fractional branes (Coulomb branch) is more difficult to
understand. We set $X^+=0$ and combine the small and large gauge
transformations in
\begin{equation}\label{sec3-gauge-sym2}
    X^-(\sigma)\cong X^-(\sigma)+\nu\tilde{\Lambda}(\sigma) X(\sigma) \csp 1 \int^{2\pi}_0\frac{d\sigma}{2\pi}\tilde{\Lambda}\in\mathbb{Z}
\end{equation}
Where $\tilde{\Lambda}$ is a periodic function.

For $X(\sigma)\neq0$~ we choose a gauge fixed solution by setting
$X^-=constant$~ identified by the shifts:
\begin{equation}\label{sec3-harmonic}
    X^-=X^-_0 \csp1 X^-\cong X^- +\nu\left(\frac1{2\pi}\int
    \frac{d\sigma}{X(\sigma)}\right)^{-1}.
\end{equation}
In addition to this $X^-$, the Coulomb branch is parameterized by
$X(\sigma)$ and $Y(\sigma)$ arbitrary functions of $\sigma$.
Therefore the solutions allow a separation of the fractional
branes both in the singular plane and in the transverse
directions. Note, however, that the gauge symmetry is completely
broken for $X\neq 0$. This is different from the fractional branes
of the \zn orbifold, which have a $U(1)$ gauge symmetry for each
fractional brane. We identify this peculiarity as arising from the
existence of a 2-dim space of orbifold fixed points.

Pursuing the analogy of T duality, we expect that there is a field
redefinition which brings the variables to the form of some gauge
fields. Indeed for $X^+\neq 0$ we define the fields
\begin{equation}\label{sec3-gauge fields}
    A\equiv \frac{X}{X^+}\hsp2 Y^+=X^+ \hsp2 Y=X^--\frac12\frac{X}{X^+}
\end{equation}
$Y^+$,$Y$ are gauge invariant and $A\cong
A+\nu\partial_\sigma\Lambda$. Of course, in the action expressed
using these variables the gauge field is decoupled from the gauge
invariant operators. In addition, it confirms the geometric
picture we have. We know of space-like cycles the null orbifold
possesses. The gauge field is exactly the coordinate which
parameterizes the space-like cycles. On the singular plane, we
encounter the same phenomenon, the field
\begin{equation}
    A^{\text{sing}}=\frac{X^-}{X}
\end{equation}
is again the natural gauge field in agreement with the existence
of null cycles.

\section{Twisted Closed String Condensation in the Null-Orbifold}\label{section4}

This section contains the main result of the paper. We show that
after condensing a closed twisted sector state, the D(-1) action flows
to that of a $\Zf_N$ orbifold (at the limit of infinite boost which we
elaborate below).

\subsection{The action after closed string condensation}

We denote by $T_k$ the modulus squared of the k twisted sector
closed string state, $k\neq 0$. This choice follows
\cite{Adams:2001sv}. We repackage the twisted condensate field
using the $\sigma$ variable as
\begin{equation}
    T(\sigma)=\sum_{k\neq 0}T_ke^{ik\sigma}
  \end{equation}
The quantum $U(1)$ symmetry of the $\Zf$ orbifold is given by
shifts in $\sigma$, and $T(\sigma)$ breaks it with the appropriate
charges. In order to flow to a $\Zf_N$ orbifold, we turn on only
$T_k$ for $k=N\cdot \Zf$, i.e, $T(\sigma)=T(\sigma+ 2\pi/N)$.

We also define the function $U(\sigma)$ to be the double integral
of $T$ which is periodic and integrates to zero
\begin{equation}
  U(\sigma)=\int\int^\sigma T(\sigma)= - \sum_{k\neq 0} \frac{T_k}{k^2} e^{ik\sigma}
\end{equation}

The couplings of the twisted condensate to open string fields are
determined uniquely from three properties. The first is locality
in the $\sigma$ variable. The second is that $T$ couples to a
quadratic form of the $X$'s. By gauge invariance the possibilities
are $X^+(\sigma)X^+(\sigma)$ and $X^2-2X^+X^-$. The third
requirement is that since the background is invariant under
translations in $X^-$, we can choose the twisted condensate field
to have a similar property.  This determines the coupling to be
\begin{equation}\label{sec4-deform}
    \int d\sigma\, T(\sigma)X^+(\sigma)X^+(\sigma)
  \end{equation}

This is also the result obtained from taking the limit of the
$\Zf_N$ orbifold. As in  \cite{Adams:2001sv}, it means that the
twisted sector closed string state couples to the open string
field which measures the effective distance from the singularity.

It is convenient to change variables to the following gauge invariant ones by
\begin{equation}
    (X^+,X^-,X)\rightarrow(X^+,L)\csp1L(\s)=2X^+(\s)X^-(\s)-X^2(\s)
\end{equation}
Using these variables the action becomes
\begin{equation}\label{sec4-uv-act}
    \mathrm{S}=\int_0^{2\pi}\left[(X^+)^2(\dot Y^a)^2
    -\frac13\frac{d(X^+)^3}{d\s}\left(\frac{d}{d\s}\left(\frac{L}{X^+}\right)+\frac2{X^+} {\dot U}(\sigma)\right)-\frac{\nu^2}{12}\left(\dot X^+\right)^4\right]
\end{equation}
The equations of motions, in terms of the gauge invariant
variables are:
\begin{subequations}
\begin{align}
    \label{sec3-deomy}
    & \frac{d}{d\sigma}\left[(X^+)^2\,\frac{d Y^a}{d\sigma}\right]=0\\
    \label{sec3-deomL}
    & \frac d{d\s}\left(X^+\frac {dX^+}{d\s}\right)+\left(\frac{dX^+}{d\s}\right)^2=0\\
    \label{sec3-deomxp}
    & X^+\left[\sum_a\left(\frac{dY^a}{d\s}\right)^2+\frac12\frac{d^2L}{d\s^2}+T(\sigma)\right]
    -\frac{d}{d\s}\left[\left(L\,\frac{d X^+}{d\s}\right)-\frac{\nu^2}{6}\left(\frac{dX^+}{d\s}\right)^3\right]=0
\end{align}
\end{subequations}
Using the fact that $X^+$ and $Y^a$ have to be constant by the
first two equations (and periodicity conditions), we can reduce
the last equation to:
\begin{equation}
    X^+\left[\frac12\frac{d^2L}{d\s^2}+T(\sigma)\right]=0
\end{equation}
and the general solution is:
\begin{equation}
  \label{sec4-cls}
    X^+(\sigma)=X^+_0\csp1 L(\sigma)=L_0-2 U(\sigma)
\end{equation}
In the next subsection we will deal carefully with the IR action
around this solution, but we can already see the glimpses of
$\Zf_N$ orbifold. Given a value of the gauge invariant $X^+_0$ and
$L_0$ we have an entire gauge orbit of the gauge symmetry
\eqref{sec3-gauge-sym}. For generic values of $X^+_0$ and $L_0$,
the gauge group is completely broken on this orbit. The maximal
unbroken gauge group occur on the orbit
\begin{equation}
  \label{lzero}
  X^+_0=0
  \csp1 L_0={\bar L}_0\equiv 2\min_{\s\in[0,2\pi]} U(\sigma)
\end{equation}
where it is $\U1^{N-1}$. This point corresponds to bringing as
many D(-1) instantons as possible close to the singularity, and
the symmetry   suggests $D(-1)$ instantons near a $\Zf_N$
singularity.

To show this we note first that for $X^+\neq0$ the gauge symmetry
is completely broken due to the transformation ~$X\rightarrow
X+\nu\pt\Lambda X^+$. For $X^+=0$, $X$ is gauge invariant and the
transformation of $X^-$ \eqref{sec3-gauge-sym2}:
\begin{equation*}
    X^-\rightarrow X^-+\nu\tilde\Lambda X=X^-+\nu\tilde\Lambda \sqrt{-L}
\end{equation*}
Note that $L\leq 0$ everywhere for $X^+=0$.  We see that if $L_0$
attains its value from \ref{lzero}, there are $N$
points\footnote{$N$ is the number of points where ~$\iint^\s
T(\s')$ reaches it's maximal value.} where $L=0$, and hence $X=0$,
and the symmetry is restored. We identify the gauge symmetry as
\begin{equation*}
    \Lambda\propto\delta(\sigma-\sigma_i)\csp1\text{for each $\sigma_i$ such
that } L(\sigma_i)=X(\sigma_i)=0
\end{equation*}
The constraint $\int\tilde\Lambda\in2\pi\mathbb{Z}$ ~removes one
transformation to obtain the gauge symmetry $U(1)^{N-1}$ of D(-1)
instantons in \zn like orbifold.

Although this is an indication, it can not be taken to be the
complete   story. This is so because
the localized transformations as we have written them are
difficult to extend to the entire Higgs branch (where the gauge
symmetry is generically broken). In this case we have to smear the
localized gauge transformations (to avoid
$\delta^2(\sigma-\sigma_0)$ terms), and there seems to be
considerable arbitrariness in how to do so.

\subsection{The "IR" theory}

Since the action is not positive definite, we first need to
clarify the notion of ``IR'' physics. By this we mean that we
separate the degrees of freedom in the action into slow and fast
variables, where for the latter we carry out a stationary phase
approximation. To obtain a clear separation of scales we write the
twisted condensate field as
\begin{equation*}
  T(\sigma)=M\cdot {\hat T}(\sigma)
\end{equation*}
and work in the scaling of ${\hat T}$ fixed and $M\rightarrow \infty$.

We are interested in working slightly off-shell - i.e, we do not
impose the equations of motion, but restrict our attention to
fluctuations which have finite action (does not scale with M). We
also would like to work near the singularity. To do so, we need to
work close to the special solution \eqref{lzero} for which the
$U(1)^{N-1}$ is unbroken. This means that we should take
\begin{equation}
  L(\sigma)+\delta L(\sigma)= \biggl( {\bar L_0}-2 U(\sigma)
  \biggr) + \delta L(\sigma)
\end{equation}
For this class $L$ scales linearly in $M$ in all of the interval
(due to the term in the parenthesis) except in the $N$ points
$\sigma_i$, where its value is held fixed as $M\rightarrow \infty$
(and determined by $\delta L$). Expanding the action to quadratic
order in $\delta L$, $\delta X^+$ and $\delta Y$ around this
solution, we obtain
\begin{equation}\label{sec4-action-fluc}
    \mathrm{S}_{fluc}=\int\frac{d\sigma}{2\pi}\left[
    (X^+_0)^2(\dot{\delta  Y^a})^2
    -(X^+_0)\frac{d(\delta X^+)}{d\s}\frac{d({\delta L})}{d\s}
    +L\left(\frac{d(\delta X^+)}{d\s}\right)^2\right]
\end{equation}

In the intervals between the points $\sigma_i$, the value of $L$
scales with $M$, and the variations $\delta X^+$ are fast in these
intervals. We therefore ``integrate out'' the variations of $X^+$
between the $\sigma_i$'s. These give the constrains that the
function $X^+(\sigma)$ is piecewise constant, and may jump at the
points $\sigma_i$. The function is therefore:
\begin{equation}
  X^+(\sigma)= X_0^+ + \delta X^+_i, \ \ \ \ \sigma_i<\sigma<\sigma_{i+1}
\end{equation}
However \eqref{sec4-action-fluc} is not convenient at the points
$\sigma_i$ since the last term is $0\times
\delta^2(\sigma-\sigma_i)$, rather we go back to
\eqref{sec4-uv-act}. Using the derivative of $X^+$ we see that the
action localizes at the points $\sigma_i$.

Next we evaluate the term $\partial_\sigma(L/X^+)-2{\dot U} /X^+$
at the points $\sigma_i$. The first step is to evaluate the
behavior of $L$ near the points $\sigma_i$. In order to obtain a
finite action we need to impose that $L/X^+$ be continuous at the
points $\sigma_i$. Otherwise we will have a
$\delta^2(\sigma-\sigma_i)$ divergence. We will use the notation
\begin{equation}
    L^\pm_i = \lim_{\sigma\rightarrow \sigma_i\pm} L(\sigma)
\end{equation}
and hence
\begin{equation}\label{sec4-dis-cond}
    \frac{L^+_i}{X^+_i}=\frac{L^-_{i}}{X^+_{i-1}}
\end{equation}

The degrees of freedom of $L$ in the intervals between the points
$\sigma_i$ do not appear in the action and can be integrated out.
To obtain some physical intuition for the remaining degrees of
freedom in $L$, we impose on L the equations of motion piecewise
in these intervals. The solution we obtain in each interval is
\begin{equation}\label{sec4-exp2}
    L(\s)=(L_i^++2
    U(\sigma_i))+\frac{L^-_{i+1}-L^+_{i}}{\s_{i+1}-\s_{i}}(\s-\s_i)-2
    U(\sigma), \ \ \ \ \s\in(\s_i,\s_{i+1})
  \end{equation}
Note that the value of $U$ is the same in all the points $\sigma_i$.

Applying the expansion \eqref{sec4-exp2} to the action
\eqref{sec4-uv-act}, (we rename the fluctuation fields by omitting
the $\delta$'s in front to unclutter the equations) we find the
leading IR action \footnote{In the calculation of
\eqref{sec4-act-ir} we used a "non-symmetrized" value to the
integration of a delta function over a discontinues functions
\begin{equation*}
    \int d\s\delta(\s-\s_i)F(\s)\equiv \lim_{\epsilon\rightarrow0}F(\s_i-\epsilon)
\end{equation*}
Any other definition of the integral will result in an equivalent
action up to a linear combination of the $L^{\pm}_i$'s.} :
\begin{multline}\label{sec4-act-ir}
    \mathrm{S}_{IR}=(X^+_0)^2\int_0^{2\pi}\frac{d\sigma}{2\pi}(\dot Y^a)^2
    -(X^+_0)^2\sum_{\s_i}\left(X^+_{i+1}-X^+_{i}\right)\frac1{(\s_{i+1}-\s_{i})}\left(\frac{L^+_{i+1}}{X^+_{i+1}}-\frac{L^+_{i}}{X^+_i}\right)
\end{multline}

In the next subsection we will match the $X^+-L$ action to that of
the boosted $\Zf_N$ singularity.

The situation of the $Y$'s is less clear. To the order that we are
working in the open string fields and in the twisted condensate
field we do not get a clear separation into $N$ degrees of freedom
below a gap. Perhaps this happens at higher orders.

\subsubsection{The boosted \zn}
We would like to clarify the term "infinitely boosted $\Zf_N$"
used in the beginning of the section. We start from the action of
a fractional $D$-brane in the \zn orbifold as computed in
\cite{Adams:2001sv}, apply infinite boost similar to the one used
in section \ref{section2} but keeping the parameter N
constant\footnote{In other words we redefine coordinate but do not
change the identification of the orbifold.} The action for a
$D(-1)$ instanton reads\footnote{We are using a slightly different
normalization for the $Z$'s then \cite{Adams:2001sv}.}:
\begin{multline}\label{sec4-zn-act}
    \mathrm{S}^{\Zf_N}=-\sum_{j=0}^{N-1}\left(X^0_{j+1,j+1}-X^0_{j,j}\right)^2\abs{Z_{j,j+1}}^2+\sum_{a=3}^9\sum_{j=0}^{N-1}\left(Y^a_{j+1,j+1}-Y^a_{j,j}\right)^2\abs{Z_{j,j+1}}^2
    +\\+\frac
    14\sum_{j=0}^{N-1}\left(\abs{Z_{j,j+1}}^2-\abs{Z_{j-1,j}}^2\right)^2
\end{multline}
Expanding the action around a general classical solution in the
Higgs branch:
\begin{align}
    & X^0_{j,j}=X^0+\delta X^0_{j,j}\\
    & Y^a_{j,j}=Y^a+\delta Y^a_{j,j}\\
    & \abs{Z_{j,j+1}}=\abs{Z}+\delta\abs{Z_{j,j+1}}
\end{align}
and dropping the $\delta$'s we find:
\begin{multline}\label{sec4-zn-act-fluc}
    \mathrm{S}^{\Zf_N}_{fluc}=-\abs{Z}^2\sum_{j=0}^{N-1}\left[\left(X^0_{j+1,j+1}-X^0_{j,j}\right)^2-\left(\abs{Z_{j,j+1}}-\abs{Z_{j-1,j}}\right)^2\right]+
    \\+\abs{Z}^2\sum_{a=3}^9\sum_{j=0}^{N-1}\left(Y^a_{j+1,j+1}-Y^a_{j,j}\right)^2
\end{multline}

We boost and rotate the coordinates  by:
\begin{equation}
    \frac1{\sqrt2}\begin{pmatrix}X^++X^-\\X^--X^+\\\sqrt2X\end{pmatrix}=
    \begin{pmatrix}\frac{\alpha}{\sqrt\beta} &\sqrt\beta\,\alpha&0 \\ \sqrt\beta\,\alpha&\frac{\alpha}{\sqrt\beta}&0\\0&0&1\end{pmatrix}
    \begin{pmatrix}X^0\\\re Z\\\im Z\end{pmatrix}\csp1
    \alpha=\frac{\sqrt\beta}{\sqrt{1-\beta^2}}
\end{equation}
Taking the limit\footnote{The explicit parametrization of the
transformation was chosen to produce a finite limit.}
$\alpha\rightarrow\infty$ we find:
\begin{gather}
    X^0=\sqrt2\alpha  X^++\frac{X^-}{2\sqrt2\alpha}+O(\alpha^{-2})\cr
    \abs{Z}^2=2\alpha^2(X^+)^2+\left((X)^2-X^+X^-\right)+O(\alpha^{-1})
\end{gather}

Plugging the transformation into \eqref{sec4-zn-act-fluc} and
taking the leading order in $\alpha$ (using only the first index
of each field).
\begin{multline}\label{sec4-act-zn-ir}
    \mathrm{S}^{\text{boosted-}\Zf_N}_{fluc}=
     -2\alpha^2(X^+_0)^2\sum_{j=0}^{N-1}
    \left(X^+_{j+1}-X^+_{j}\right)\left(\frac{L_{j+1}}{X^+_{j+1}}-\frac{L_{j}}{X^+_{j}}\right)+
    \\+2\alpha^2\abs{X^+_0}^2\sum_{a=3}^9\sum_{j=0}^{N-1}\left(Y^a_{j+1,j+1}-Y^a_{j,j}\right)^2+O(\alpha^{0})
\end{multline}
The factor $\alpha$ can be absorbed into a rescaling of the
coordinates.

\subsubsection{Relating the IR physics}
We found the action of the IR physics around a Higgs branch VEV
both in the boosted-$\Zf_N$ \eqref{sec4-act-zn-ir} and the
null-orbifold \eqref{sec4-act-ir}. Focusing on the orbifold
directions:
\begin{align*}
    \mathrm{S}^{\text{boosted-}\Zf_N}_{IR}&\supset
     -(X^+_0)^2\sum_{j=0}^{N-1}\left(X^+_{i+1}-X^+_{i}\right)
    \left(\frac{L_{i+1}}{X^+_{i+1}}-\frac{L_{i}}{X^+_i}\right)\\
    \mathrm{S}_{IR}&\supset-(X^+_0)^2\sum_{\s_i}\frac1{(\s_{i+1}-\s_{i})}\left(X^+_{i+1}-X^+_{i}\right)
    \left(\frac{L^+_{i+1}}{X^+_{i+1}}-\frac{L^+_{i}}{X^+_i}\right)
\end{align*}
The above actions are exactly the same up to rescaling of fields.  The
action is ill-defined at the singularity, (i.e the VEV of the $X^+$
field vanish) which indicates the emergence of new degrees of freedom,
which are the fractional branes (Coulomb branch) together with the
expected $\U1^{N-1}$ gauge symmetry.

Note, however, that after the rescaling of the fields, the issue
of whether one is dealing in a  finite boost or an infinite boost
is subleading in the boost parameter. In order to see this effect
in the flow from the null orbifold to the \zn case, we need to
look at subleading corrections to the action - in this case
subleading in $M$. We do not know how to do this precisely, and
hence can not answer the detailed question of whether the deformed
null singularity is the infinitely boosted \zn or boosted by some
parameter proportional to a power of $M$.

Note also that we cannot make a clear study of the IR physics in the
Coulomb branch or the $Y$ variables with our probes as already pointed
out.

\section{Summary and Conclusions}\label{section5}

The purpose of this paper is to study the geometry of small
deformations of the null orbifold, which occur after condensation
of twisted sector states. This might be a first step towards
understanding both the situation in which twisted sector states
are condensed with large VEV's, or the situation of pair creation
of twisted sector state, which might be ways in in which the
singularity might be tamed.

We focused on the case of the null-orbifold since it has a clear
relationship to the \zn orbifold - the latter is the large N limit
of the former. We have exhibited, using D(-1) brane probes, some
evidence that indeed there is a transition from the null orbifold
to the \zn orbifold.

In the case of flows between \zn theories, one can eventually flow
to flat space, smoothing out the singularity completely. We expect
that a similar situation occurs here - hence we conclude that the
null orbifold may be smoothed out by the condensation of twisted
sector states.

The intermediate step - of flowing to a \zn - might be interesting
by itself. For example, it might be interesting to study the
condensation of twisted sector states of the BTZ black hole and
examine their relation of the deformed geometry to the microstates
of \cite{Mathur:2005zp}.

\acknowledgments

The  authors are happy to thank O.Aharony, D.Kutasov, B.Pioline,
S.Ross, M.Rozali, J.Simon and S.Shenker  for useful discussions. The
work is supported in part by the Israel Science Foundation, by the
Braun-Roger-Siegl foundation, by the European network
HPRN-CT-2000-00122, by a grant from the G.I.F. (the German-Israeli
Foundation for Scientific Research and Development), by the Minerva
Foundation, by the Einstein Center for Theoretical Physics and the
by Blumenstein foundation.

\appendix

\section{Calculations of the Wave-Functions Limit}\label{appendix a}

In this appendix we demonstrate (in the untwisted case) the
limiting procedure between wave-functions on the \zn orbifold and
wave-functions on the null-orbifold. We start with the
wave-functions of the \zn orbifold in the static coordinates
\eqref{sec2-zn-wf}
\begin{equation}\label{appa-wf-zn}
    \Psi_{k,l}^{N}=N e^{-ik^0x^0+iN\phi
    l}J_{Nl}(2u)\csp1
    u{e^{i\phi}}\equiv-i\bar kz
\end{equation}
The limit procedure is defined in \eqref{sec2-wf-limit} and
\eqref{sec2-symmetric}, the boost matrix is
\begin{equation}
    M_N=\begin{pmatrix}\frac {a}{\sqrt{a^2-b^2}}&0& \frac {b}{\sqrt{a^2-b^2}}\\
                       \frac {b}{\sqrt{a^2-b^2}} & 0 &\frac {a}{\sqrt{a^2-b^2}}\\
                       0 & -1 & 0 \end{pmatrix},
    \hsp3\text{with}\qquad
    a,b=\frac{\sqrt{1\pm 1/N^2}}{\sqrt2}
\end{equation}
We take care of each term of \eqref{appa-wf-zn} separately. First
the phase factors:
\begin{align}
    e^{-ik^0x^0}\xrightarrow{\text{boost}}\,&
    e^{iN^2(ak_0+bk_1)(ax^0-bx^1)}=\cr
    &=\exp\left[\frac{i}{2}N^2(k_0+k_1)(x^0-x^1)+\frac{i}{2}k_0x^0+\frac{i}{2}k_1x^1+O(1/N)\right]
\end{align}
\begin{align}
    e^{iN\phi l}=&\left(-\frac{k\bar{z}}{\bar
    kz}\right)^{-\frac{Nl}2}\xrightarrow{\text{boost}}\
    \left[-\frac{\left(\frac{i(ak_1+bk_0)}{\sqrt{a^2-b^2}}-k_2\right)\left(\frac{i(bx^0-ax^1)}{\sqrt{a^2-b^2}}-x^2\right)}{c.c}\right]^{-\frac{Nl}2}=\cr
    &
    =(-1)^{\frac{Nl}2}\left[\frac{-(k_0+k_1)(x^0-x^1)-i\sqrt2~\frac{x^2(k_0+k_1)+k_2(x^0-x^1)}{N}+O(1/N^2)}{c.c}\right]^{-\frac{Nl}2}\cr
\end{align}
The large N limit is easily calculated via $(1+x/N)^N\rightarrow
e^x$ and reduces to
\begin{equation}
    e^{iN\phi l}\xrightarrow{\text{boost}}\,
    (-1)^{\frac{Nl}2}\exp\left[-il\sqrt{2}~\frac{x^2(k_0+k_1)+k_2(x^0-x^1)}{(x^0-x^1)(k_1+k_0)}+O(1/N)\right].
\end{equation}
The remaining part of the story is the Bessel function. We quote
the asymptotic expansion of the Bessel function from \cite{I.S
Gradshteyn I.M Ryzhik:1971} which shall play a major role
\begin{align}\label{GR-formula}
    J_\nu(z)=&\sqrt{\frac{2}{\pi
    z}}cos(z-\frac{\pi}{2}\nu-\frac{\pi}{4})\left[\sum^{n-1}_{k=0}\frac{(-1)^k}{(2z)^{2k}}\frac{\Gamma(\nu+2k+\frac{1}{2})}{(2k)!\Gamma(\nu-2k+\frac{1}{2})}+R_1\right]-\cr
    &- \sqrt{\frac{2}{\pi z}}sin(z-\frac{\pi}{2}\nu-\frac{\pi}{4})
    \left[\sum^{n-1}_{k=0}\frac{(-1)^k}{(2z)^{2k+1}}\frac{\Gamma(\nu+2k+\frac{3}{2})}{(2k+1)!\Gamma(\nu-2k-\frac{1}{2})}+R_2\right]\cr
\end{align}
Where for $n>\nu/2-1/4$ the remainders satisfy
\begin{equation}
    \left|R_1\right|<\left|\frac{\Gamma(\nu+2n+1/2)}{(2z)^{2n}(2n)!\Gamma(\nu-2n+1/2)}\right|
\end{equation}
\begin{equation}
    \left|R_2\right|<\left|\frac{\Gamma(\nu+2n+3/2)}{(2z)^{2n+1}(2n+1)!\Gamma(\nu-2n-1/2)}\right|
\end{equation}
The Bessel functions in \eqref{appa-wf-zn} transforms under the
boost:
\begin{align}
    J_{Nl}&\left[\sqrt{\left((k^1)^2+(k^2)^2\right)\left((x^1)^2+(x^2)^2\right)}\right]\xrightarrow{\text{boost}}\cr
    =&J_{Nl}\left[N^2\abs{x^+k^+}+\frac{(k^2x^+)^2-k^+k^-(x^+)^2-(k^+)^2x^+x^-+(k^+x^2)^2}{2\abs{x^+k^+}}+O(1/N)\right]\cr
\end{align}
Using the useful limit
\begin{equation*}
    \lim_{z\rightarrow\infty}\frac{\Gamma(z+a)}{\Gamma(z)}z^{-a}=1
\end{equation*}
We observe that the terms in the Bessel function expansion don't
scale as powers of N but only of n and hence no terms can be
dropped. Fortunately, we are able to re-sum the series
\eqref{GR-formula} in the $N\rightarrow\infty$ limit. As
emphasized in \eqref{GR-formula} we should carefully estimate the
remainder. Let $\nu=Nl$ then we take $n$ such that (at least)
$\nu=2n$ and also $z=\frac{\alpha}{2} n^2$ for some constant
$\alpha$. Estimating the remainder we obtain
\begin{equation*}
    \left|\frac{\Gamma(\nu+2n+1/2)}{(2z)^{2n}(2n)!\Gamma(\nu-2n+1/2)}\right|=\left|\frac{\Gamma(4n+1/2)}{(\alpha
    n^2)^{2n}(2n)!\Gamma(1/2)}\right|<\frac{1}{\alpha^{2n}}\frac{(4n)^{2n}}{(n^2)^{2n}}=(\frac{4}{\alpha})^{2n}(\frac{1}{n})^{2n}
\end{equation*}
These are preferable circumstances and it holds that
\begin{equation*}
    \lim_{k\rightarrow\infty}\sum_{l=0}^{k}P(l,k)=\sum_{l=0}^{\infty}P(l,\infty)
\end{equation*}
Which means that
\begin{multline}
    \lim_{N\rightarrow\infty}J_{Nl}(N^2\abs{x^+k^+}+A)
    =\lim_{N\rightarrow\infty}\sqrt{\frac{2}{\pi
    N^2\abs{x^+k^+}}}\cos\left(N^2\abs{x^+k^+}+A-\frac{\pi}{2}Nl-\frac{\pi}{4}\right)\\
    \cdot\left[\sum^{\left\lceil
    Nl/2\right\rceil-1}_{k=0}\frac{(-1)^k}{(2N^2x^+k^+)^{2k}} \frac{\Gamma(Nl+2k+\frac{1}{2})}{(2k)!\Gamma(Nl-2k+\frac{1}{2})}+R_1\right]
    -\text{$2^{nd}$ term}
\end{multline}
where
\begin{equation*}
    A\equiv \frac{(k^2x^+)^2-k^+k^-(x^+)^2-(k^+)^2x^+x^-+(k^+x^2)^2}{2\abs{x^+k^+}}+O(1/N)
\end{equation*}
 Taking the limit inside the square brackets gives
\begin{align}
    =&\left[\lim_{N\rightarrow\infty}\sqrt{\frac{2}{\pi
    N^2\abs{x^+k^+}}}\cos\left(N^2\abs{x^+k^+}+A-\frac{\pi}{2}Nl-\frac{\pi}{4}\right)\right]
    \left[\sum^{\infty}_{k=0}\frac{(-1)^kl^{4k}}{(2k)!(2x^+k^+)^{2k}}\right]-
    \text{$2^{nd}$ term}\cr
    =&\left[\lim_{N\rightarrow\infty}\sqrt{\frac{2}{\pi
    N^2x^+k^+}}\cos(N^2x^+k^++A-\frac{\pi}{2}Nl-\frac{\pi}{4})\right]\cos\left(\frac{l^2}{2\abs{x^+k^+}}\right)-\cr
    &- \left[\lim_{N\rightarrow\infty}\sqrt{\frac{2}{\pi
    N^2\abs{x^+k^+}}}\sin\left(N^2\abs{x^+k^+}+A-\frac{\pi}{2}Nl-\frac{\pi}{4}\right)\right]\sin\left(\frac{l^2}{2\abs{x^+k^+}}\right)\cr
    =&\lim_{N\rightarrow\infty}\sqrt{\frac{2}{\pi  N^2|x^+k^+|}}\cos\left(N^2\abs{x^+k^+}+A-\frac{\pi}{2}Nl-\frac{\pi}{4}+\frac{l^2}{2\abs{x^+k^+}}\right)
\end{align}
Combining the pieces of the full wave function together
\begin{align}
   \Psi^{\text{boosted}}_{k;\,l}&(x^+,x^-,x^2)=
    \sqrt{\frac{2}{\pi\abs{x^+k^+}}}e^{\pi i\frac{Nl}2}
    e^{-il\frac{x^2k^+ -x^+k_2}{x^+k^+}}
    e^{\left[-i{N^2}k^+x^+-\frac{i}{2}\left(k^+x^-+k^-x^+\right)+O(1/N)\right]}\cr
    &\cdot\cos\biggl[N^2\abs{x^+k^+}-\frac{\pi}{2}Nl-\frac{\pi}{4}+\frac{l^2}{2\abs{x^+k^+}}+\cr
    &\qquad
    +\frac{(k^2x^+)^2-k^+k^-(x^+)^2-(k^+)^2x^+x^-+(k^+x^2)^2}{2\abs{x^+k^+}}+O(1/N)\biggr]=\cr
    =&
    \frac{(\text{sign}(k^+x^+))^{Nl}}{\sqrt{2\pi ix^+k^+}}
    \Biggl[
    e^{\biggl[i\frac{\left(l-x^2k^++x^+k^2\right)^2}{2x^+k^+}+ik^2x^2-ik^+x^--ik^-x^++O(N^{-1})\biggr]}+\cr
    &+(-)^{Nl+1}e^{\biggl[-i\frac{\left(l+x^2k^+
    -x^+k^2\right)^2}{2x^+k^+}-ik^2x^2-i2N^2k^+x^++O(N^{-1})\biggr]}
    \Biggr]
\end{align}
This is the expression (up to trivial algebra) quoted in the text
\eqref{sec2-result of boost}. The First exponential is the wave
function of the null-orbifold and the second produces an
expression which is interpreted as a vanishing distribution. In
order to prove the last statement we study the integration of the
second exponential in the boosted wave-function with a test
function $g(x^+,x)$ at the large $N$ limit:
\begin{equation*}
    \int dx^+dx^2 g(x^+,x^2)\frac 1 {\sqrt{2\pi ix^+k^+}}\exp\biggl[-i\frac{\left(l+x^2k^+\right)^2}{2x^+k^+}-i2{N^2}k^+x^+\biggr]
\end{equation*}
Without loss of generality we substituted $k^2=0$. The  $x^+$
integration can now be evaluated using a saddle point method:
\begin{align}\label{appa-sad-point}
   \int dx^+dx^2 & g(x^+,x^2)\frac 1 {\sqrt{2\pi ix^+k^+}}\exp\biggl[-i\frac{\left(l+x^2k^+\right)^2}{2x^+k^+}-i2{N^2}k^+x^+\biggr]=\cr
    \propto&\sum_{\pm}\int dx^2\frac1{Nk^+}g\left(\pm\frac{(l+x^2k^+)}{2Nk^+},x^2\right)
    e^{\mp 2iN(l+x^2k^+)}\cr
    \propto&\sum_{\pm}\frac {\wt G_{\pm}^N(\pm
    2Nk^+)}N\xrightarrow{N\rightarrow\infty} 0
\end{align}
Where $\wt{G}_{\pm}^N$ is the Fourier transform (with respect to
$x^2$) of:
\begin{equation*}
    G_{\pm}^N\equiv g\left(\pm\frac{(l+x^2k^+)}{2Nk^+},x^2\right)
\end{equation*}
For a large class of functions  \footnote{We didn't carry a full
classification of the functions $g(x^+,x^2)$ that have the above
property. A large enough set of examples are polynomials in
$x^+,x^2$ multiplied by decaying exponentials and Gaussian}
$g(x^+,x^2)$ (which are in particular $\mathbb{L}_2$) the limit
\eqref{appa-sad-point} is indeed zero which proves our claim.

\section{First Quantization of the String on $\Rf^{1,2}/\Gamma$}\label{appendix b}

The orbifolds in mind are defined by a flat space CFT with a
quotient by a twist:
\begin{gather}
    ds^2=-d(x^0)^2+d(x^1)^2+d(x^2)^2+dx_{\parallel}^2\cr
    X\cong e^{2\pi\hat J}X={\left[e^{2\pi\J}\right]^{\mu}}_{\nu}X^\nu \csp1
    \J\in\SO{1,2}
\end{gather}
We discuss 3 cases according the 3 classes of SO(1,2):
\begin{itemize}
    \item The Milne orbifold (J is hyperbolic):
          $        J_\Delta=2\pi\Delta \hat J_{02} $
    \item The Null-orbifold (J is parabolic):
          $        J_v=\frac {2\pi v}{\sqrt2}\left(J_{02}+J_{12}\right) $
    \item The $\Zf_N$ orbifold (J is elliptic):
          $        J_N=\frac{2\pi}{N}J_{12} $
\end{itemize}
The operators above can be defined using the matrices in
$(x^0,x^1,x^2)$ basis:
\begin{equation*}
            \J_\Delta=\Delta\begin{pmatrix}0&0&1\\0&0&0\\1&0&0\end{pmatrix}\qquad
            \J_v=\frac{v}{\sqrt2}\begin{pmatrix}0&0&1\\0&0&-1\\1&1&0\end{pmatrix}\qquad
            \J_N=\frac1{N}\begin{pmatrix}0&0&0\\0&0&1\\0&-1&0\end{pmatrix}
\end{equation*}
The worldsheet action is:
\begin{equation}
    S=\frac1{4\pi}\int d^2z\,
    \frac2\alp\eta_{\mu\nu}\left(\pt X^\mu\bpr X^\nu+\eta_{\mu\nu}\psi^\mu\bpr\psi^\nu
    +\eta_{\mu\nu}\tilde\psi^\mu\pt\tilde\psi^\nu\right)
\end{equation}
And the monodromies \footnote{Remember~
$z=e^{i(\s+\tau)}~,~\bz=e^{-i(\s-\tau)}$.}:
\begin{gather}
    X(\s+2\pi,\tau)=e^{2\pi w \J}X(\s,\tau)\cr
    \psi\left(ze^{2\pi i}\right)=e^{-2\pi i\left((\nu-\frac12)\idn+iw
       \J\right)}\psi(z)\cr
    \tilde\psi\left(\bz e^{-2\pi i}\right)=e^{2\pi i\left((\tilde\nu-\frac12)\idn-iw \J\right)}\tilde\psi(\bz)
\end{gather}
where $\nu$ and $\tilde \nu$ take the values of $0$ for R-sector
and $\frac12$ for NS-sector and $~w\in\Zf~$ is the twisted sector
number.

The bosonic part mode expansion in analogy to
\cite{Pioline:2003bs} ($w\neq0$):
\begin{multline}
    \sqrt{\frac2\alp}X^\mu(\s,\tau)=
    \sqrt{\frac2\alp}\left[e^{w\J\s}\right]^{\mu}_{\rho}X_z^\rho(\alpha_{0},\tilde\alpha_{0};\tau)+\cr
    +i\sum_{n\neq0}\left[\frac{e^{-i(n+iw\J)(\s+\tau)}}{n+iw\J}\right]^{\mu}_{\rho}\alpha_{n}^\rho+
    i\sum_{n\neq0}\left[\frac{e^{i(n-iw\J)(\s-\tau)}}{n-iw\J}\right]^{\mu}_{\rho}\tilde\alpha_{n}^\rho
\end{multline}
The zero-mode part is 
\begin{subequations}
 \begin{align}
    &\text{Null}\csp{1.1}{X_z}^\mu(\tau)
    ={{\cosh\bigl(w\tau
    \J\bigr)}^\mu}_\nu x^\mu+{{\Bigl[\idn+\frac12\cosh\bigl(w\tau \J\bigr)\Bigr]}^\mu}_\nu \frac{2\alp\tau}{3}
    p^\nu\\
    &\text{Other}\csp{.8}{X_z}^\mu(\tau)
    ={{\cosh\bigl(w\tau
    \J\bigr)}^\mu}_\nu x^\mu+\left[(w\J)^{-1}\sinh(w\J\tau)\right]^\mu_\nu
    \alp p^\nu
\end{align}
\end{subequations}
Using the Euclidean world-sheet, the left/right moving parts can
be expanded:
\begin{subequations}
 \begin{gather}
    \pt X(z)=-iz^{\,-(1+iw\J)}(w\J x+\alp p)
    -i\sqrt{\frac\alp2}\sum_{n\neq0}z^{\,-(n+1+iw\J)}\alpha_n\\
    \bpr X(\bz)=i\bz^{\,-(1+iw\J)}(w\J x-\alp p)
    +i\sqrt{\frac\alp2}\sum_{n\neq0}\bz^{\,-(n+1-iw\J)}\tilde\alpha_n\\
    \psi^\mu(z)=\sum_{r\in\Zf+\nu}\left[\frac{1}{z^{r+\frac12+iw\J}}\right]^\mu_\rho\psi_r^\rho\\
    \tilde\psi^\mu(\bz)=\sum_{s\in\Zf+\tilde\nu}{\left[\frac{1}{\bz^{s+\frac12-iw\J}}\right]^\mu}_\rho\tilde\psi_s^\rho
\end{gather}
\end{subequations}
The commutation relations are calculated form the above expansions
using the OPE and canonical quantization relations (needed for the
quasi-zero modes).
\begin{align}
    &[\tilde\alpha_n^\mu,\tilde\alpha_m^\nu]=\delta_{n+m}(n\eta-iw\J\eta)^{\mu\nu}&
    &\left\{\psi^\mu_r,\psi^\nu_{r'}\right\}=\delta_{r+r',0}\eta^{\mu\nu}\cr
    &[\alpha_n^\mu,\alpha_m^\nu]=\delta_{n+m}(n\eta+iw\J\eta)^{\mu\nu}&
    &\left\{\tilde\psi^\mu_s,\tilde\psi^\nu_{s'}\right\}=\delta_{s+s',0}\eta^{\mu\nu}
\end{align}
Where we adopted the following definition of quasi-zero
modes\footnote{Note that the above definition do not consist of a
full set of zero-mode operators. This is similar to flat space
where
$\alpha_0^\mu=\tilde\alpha_0^\mu=\sqrt\frac{\alp}{2}p^\mu$.}:
\begin{equation}
    \alpha_0=\frac{\alp p+w\J x}{\sqrt{2\alp}}\qquad    \tilde\alpha_0=\frac{\alp  p-w\J x}{\sqrt{2\alp}}
    \qquad     [\alpha_0,\tilde\alpha_0]=0
\end{equation}
The Virasoro generators (matter part) are
\begin{subequations}
 \begin{align}
    &L^\psi_n 
    =\frac14\sum_{r\in\Zf+\nu}
    \left[(2r-n)\eta+iw\left(\eta\J-\J^T\eta\right)\right]_{\mu\nu}\acno{\psi^\mu_{n-r}\psi^\nu_{r}}+a^\psi(w,\nu)\delta_{n,0}\\
    &L^x_n 
    =\frac12\sum_{m}\acno{\alpha_m\eta\alpha_{n-m}}+\delta_{n,0}a^x(w)
\end{align}
\end{subequations}
Using $\left[L_1,L_{-1}\right]=2L_0$ we find:
\begin{subequations}
\begin{align}
   & a^x=\frac{w^2}4\tr(\J^2)+\bok{vac}{\frac12\eta_{\mu\rho}\alpha_{0}^\mu\alpha_{0}^\rho}{vac}_{w}\\
   & a^\psi_{NS}=-\frac{w^2}{4}\tr(\J^2)\\
   & a^\psi_{R}=\frac{D}{16}-\frac{w^2}{4}\tr(\J^2)+\bok{vac}{\left(\frac{iw}{2}\eta\J\right)_{\mu\s}\psi_{0}^\mu\psi_{0}^\s}{vac}_{w,R}
\end{align}
\end{subequations}
In order not to break worldsheet supersymmetry the quantization
scheme for the bosonic zero modes and the R-sector fermionic zero
modes must obey
\begin{equation*}
    a^\psi_{R}+a^x=\frac{D}{16}
\end{equation*}
Applying this constraint we can fix the quantization scheme for
the different classes of orbifolds (using the already discussed
scheme of the bosonic part). The zero-point energies can then be
calculated according to:
\begin{itemize}
    \item \textbf{Elliptic Orbifolds Class}: Find d+1 vectors that
    diagonalize the matrix $~iw\J\eta$. Vectors corresponding to
    positive eigenvalue $\lambda$  annihilate the vacuum. Vectors
    corresponding to zero eigenvalue are 'standard' zero-modes
    which annihilate the vacuum (these are momentum
    operators). The bosonic
    zero point energy have a contribution from negative eigenvalue
    modes, $\frac12\lambda_i$  for each negative eigenvalue (the R-sector will have the opposite
    contribution).
    \item \textbf{Hyperbolic Orbifolds Class}: Find d+1 vectors that
    diagonalize the matrix $~iw\J\eta$, their eigenvalues are pure
    imaginary. Vectors
    corresponding to zero eigenvalue are 'standard' zero-modes
    which annihilate the vacuum (actually these are momentum
    operators). Vector corresponding to (non-vanishing) pure
    imaginary zero mode should be treated as in
    \cite{Pioline:2003bs} to eliminate imaginary contribution to the zero point
    energies.
    \item \textbf{Parabolic Orbifolds Class}: Either by taking
    the limit from the hyperbolic or elliptic cases we set the
    zero-point energies to zero (the orbifold part).
\end{itemize}
The different zero-point energies are calculated for the
representatives of the classes ($\Zf_N$ , Null and Milne
orbifolds):

\begin{tabular}{|l||c|c|c|c|c|}
  \hline
          & $a^X$                    & $a^\psi_{NS}$             & $a^\psi_{R }$                     & $a^{g}_{NS}$ & $a^{g}_{R}$ \\
  \hline\hline
  Flat    & $0$                      &  $ 0 $                    &  $D/{16}$                         &  $-1/2 $     &  $-5/8 $\\
  \hline
  $\Zf_n$ & ${w}/{2N}(1-w/N) $       &  ${w^2}/{2N^2} $          &  $D/{16}-{w}/{2N}(1-w/N)$         &   $-1/2 $    &  $-5/8 $\\
  \hline
  Milne   & $\frac{w^2\Delta^2}{2}$&  $-\frac{w^2\Delta^2}{2}$ &  $D/{16}-\frac{w^2\Delta^2}{2}$ &  $-1/2 $     &  $-5/8 $\\
  \hline
  Null    & $0$                      &  $ 0 $                  &  $D/{16}$                       &  $-1/2 $     &  $-5/8 $\\
  \hline
\end{tabular}

\subsection{The Vacuum Structure}
By a simple manipulation of commutation relations we can rewrite
$L_0$ as:
\begin{equation}
    L_0
       =L_0^{(\text{diag})}
       +\frac{w}{2}(\eta\J)_{\mu\nu}\Sigma^{\nu\mu}-A
\end{equation}
With the Lorentz generators defined as
\begin{equation*}
    \Sigma^{\mu\nu}\equiv-\frac
    i2\sum_{r\in\Zf+\nu}\left[\psi^\mu_{r},\psi^\nu_{-r}\right]\csp2
    \frac{w}{2}(\eta\J)_{\mu\nu}\left[\Sigma^{\nu\mu},\psi^\s_r\right]=
    -iw{\J^{\s}}_{\mu}\psi^{\mu}_r
\end{equation*}
The diagonal part of $L_0$:
\begin{equation*}
    L_0^{(\text{diag})}=\frac12\sum_{m}\acno{\alpha_{-m}^\mu{\alpha_m}_{\mu}}
    +\frac12\sum_{r\in\Zf+\nu}r\acno{\psi_{-r}^\mu{\psi_r}_{\mu}}
    +L_0^{\text{ghost}}+a(w,\nu)
\end{equation*}
And the constant A defined as:
\begin{equation*}
    A(w,\nu)=\bok{vac}{\left(\frac{iw}{2}\eta\J\right)_{\mu\s}\psi_{-\nu}^\mu\psi_{\nu}^\s}{vac}_{w,\nu}
\end{equation*}
The constant A rises from "undoing" the normal ordering of the
$\psi$'s. It takes a non vanishing value (equal to $\frac{w}{2N}$)
only in the R-sector of the elliptic orbifold where there exist
fermion zero modes \footnote{In the Elliptic case in the sector
$w=\frac{N}{2}$ there are fermion zero modes in the NS sector, we
ignore that subtlety which is not important for the out
discussion.} (and not only quasi-zero modes).

Remembering that a physical state in the CFT must be a zero
eigenstate of $L_0$ we study the effect of the operator $~\frac
w2(\eta\J)_{\mu\nu}\Sigma^{\mu\nu}~$ on the spectrum.
\begin{itemize}
    \item \textbf{Elliptic Orbifolds Class}: The operator $~\frac
    w2(\eta\J)_{\mu\nu}\Sigma^{\mu\nu}~$ is a generator of a rotations
    group $\SO2$ in the direction of the orbifold and has real
    eigenvalues. By changing the charges of the left moving and
    right moving sides it is possible to generate physical states
    obeying $L_0=\wt L_0=0$ of "mixed" type $(\text{NS},\text{R})$.

    As is well known from the literature
    \cite{Adams:2001sv,Harvey:2001wm}, in these theories if $N$
    is an even integer, untwisted fermions cannot be introduced. By taking $N$ odd one is able to construct
    type II theory where the untwisted tachyon is projected out. The
    physical condition $L_0=\wt L_0=0$ forces us to consider
    $(\text{NS-},\text{R})$ sectors for odd $\omega$ and there are tachyons in all twisted
    sectors.
    \item \textbf{Hyperbolic Orbifolds Class}: The operator $~\frac
    w2(\eta\J)_{\mu\nu}\Sigma^{\mu\nu}~$ is a generator of a boost
    group $\SO{1,1}$ in the direction of the orbifold it has pure imaginary eigenvalues.
    These eigenvalues can be compensated by a suitable bosonic
    wave-function.
    \item \textbf{Parabolic Orbifolds Class}: The operator $~\frac
    w2(\eta\J)_{\mu\nu}\Sigma^{\mu\nu}~$ is a generator of a null-boost
    group in the direction of the orbifold. The operator is nilpotent such
    that $L_0$ can be written in a Jordan form. Thus the CFT is
    logarithmic, at this point we cannot determine whether it is
    possible to find a consistent set of constraints on the
    spectrum (i.e BRST + GSO + orbifold projection) such that the resulting theory will
    have no branch cuts (mutual locality between operators) and will be
    modular invariant.
\end{itemize}

\bibliographystyle{plain}

\end{document}